\documentclass[11pt]{article}

\usepackage[utf8]{inputenc}
\usepackage[T1]{fontenc}
\usepackage[margin=1in]{geometry}
\usepackage{graphicx}
\usepackage{booktabs}
\usepackage{tabularx}
\usepackage{amsmath}
\usepackage{amssymb}
\usepackage{textcomp}
\usepackage{microtype}
\usepackage{enumitem}
\usepackage[numbers,sort&compress]{natbib}
\usepackage[hidelinks]{hyperref}
\usepackage{url}
\usepackage{caption}
\newcommand{\antah}{\texorpdfstring{Anta\d{h}kara\d{n}a}{Antahkarana}}
\DeclareRobustCommand{\inv}[1]{\textsc{inv}\text{-}#1}
\DeclareUrlCommand\code{\urlstyle{tt}}
\makeatletter\g@addto@macro\UrlBreaks{\do\_}\makeatother

\graphicspath{{figs/}}
\title{\textbf{Falsifiable Release Gates for Self-Improving Systems:\\Standing Invariants at Scale}}
\author{Deepak Soni\\[2pt]\normalsize AI Architect\\ \normalsize \texttt{deepak.satna@gmail.com}}
\date{July 2026}

\begin{document}
\maketitle

\begin{abstract}
\noindent Safety claims for self-improving agent runtimes are almost always self-graded: a policy file, a
guardrail, a promise in a README. We describe \emph{falsifiable release gates}, a methodology in which every new
capability must pass a pre-declared, machine-checkable acceptance suite before it ships, while a fixed set of
standing invariants is preserved across every gate. We instantiate it in \antah{}, an open runtime, then do what
a method paper is only vindicated by: we follow the same runtime as it grows and ask whether the guarantees
survive. The safety-critical property, that no action reaches an effector without a capability token minted by a
control ring, is machine-checked exhaustively over the reachable states of a bounded model; a deliberately broken
model yields the shortest counterexample, so the checker demonstrably has teeth. We then carry the runtime
through six further releases. Across every one, the action-safety invariants \inv{1} through \inv{6} held without
a single change, and one release added three capabilities while introducing no new invariant. Under the same
teeth discipline, six more machine-checked families were added: memory with provable unlearning, a governed
agent, calibrated abstention over a post-quantum record, a harness of many sub-agents, the self-improvement loop
itself, and the residency of what it produces. The acceptance suite grew from 122 tests to 563. The load-bearing
result sits in the negative space: across more than a doubling of capability, the safety core was neither weakened
nor redesigned. The last families are the first on real hardware: gated self-improvement compounds a small model
from 20\% to 70\% accuracy while auto-rejecting a candidate that only inflates confidence, and the whole governed
path costs 0.021 ms per request, 0.008\% of model inference. We release the runtime, tools, and gate suite; every
number reproduces with a single command.
\end{abstract}

\section{Introduction}
\label{sec:intro}

An agent capable of altering its own behavior is useful because it is not static, and dangerous for the same
reason. The hard question is not ``is this one version safe?'', but ``does every new capability preserve the
safety of the last?'' as agent runtimes gain memory, tool access, multi-agent coordination, and the ability to
revise their own policies~\citep{yao2023react,schick2023toolformer,wang2024survey}. Today that question is
answered, if at all, by inspection and good intentions~\citep{amodei2016concrete}. The system's authors say that
a guardrail exists, and the evidence for that is that it is self-graded for safety.

We contend that self-improving systems need a process for safety, not a one-off audit, and that the process can
be made falsifiable. Our suggestion is the falsifiable release gate. Every capability the system acquires ships
behind a gate: a pre-declared, machine-checkable acceptance suite that must pass before the code is considered to
exist. A small set of standing invariants is maintained across all the gates, so later capabilities cannot
undermine the guarantees of earlier ones. And the safety-critical core is machine-checked (verified on a model)
instead of asserted.

\begin{figure}[htbp]
\centering
\includegraphics[width=0.86\linewidth]{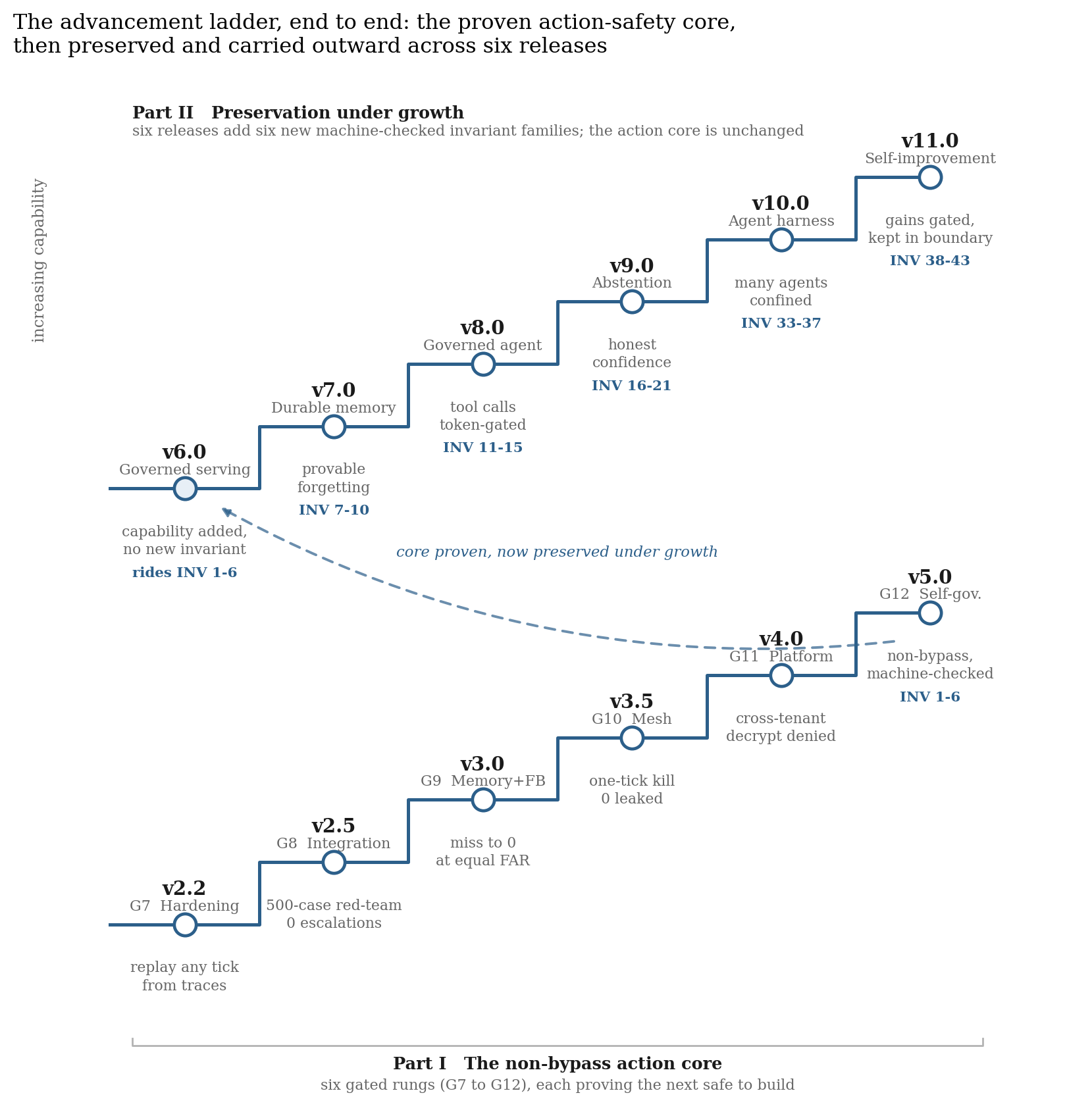}
\caption{The advancement ladder, end to end. Part I (lower tier) is the proven action-safety core: six
gated rungs from G7 to G12, each proving the next safe to build, with the non-bypass property machine-checked
at G12 (\inv{1} to \inv{6}). Part II (upper tier) carries that same core through six further releases; each
rung names the invariant family it introduced, and governed serving (v6.0) is marked as adding capability while
introducing no new standing invariant. The core is preserved, unchanged, across the whole climb.}
\label{fig:ladder}
\end{figure}

We concretely develop the method by building a runtime, \antah{}, up a ladder of six gates
(Figure~\ref{fig:ladder}), from the ability to reconstruct any past decision from traces (G7) through adversarial
tool-use integration (G8), drift-free learning (G9), fleet-scale governance (G10), multi-tenant isolation (G11),
and finally a self-governing loop whose non-bypass property is machine-checked and whose self-modification is
contained by construction (G12). Each rung is chosen to make the next one safe to build. One does not attempt a
self-improving loop before one can prove control over a fleet in a single tick. One does not attempt that before
the gate can learn without loosening itself.

A method shown on a system frozen at the moment of demonstration, though, has only been shown to work once. The
interesting claim is not that the invariants held once, at the moment the ladder was first built. It is whether
they keep holding while the runtime actually changes underneath them. So the second half of this paper is the
evidence for the durable form of the claim. We take the same open runtime and run the method forward through six
more releases, each a real body of work rather than a point release: governed model serving with nightly
consolidation, then durable governed memory with provable forgetting, then a governed agent that uses tools in
parallel, then calibrated abstention backed by a post-quantum audit trail, then a long running agent harness that
runs many governed sub-agents under one control ring, and finally the loop that changes the system itself, gated
self-improvement whose adoptions we then keep inside the boundary that owns them. At each release we re-ran the
invariant checkers and recorded two things. First, whether the original action-safety invariants still held
without modification. Second, what new invariants the release added, and whether each of them came with the
broken models that prove the checker can fail.

The contribution is the method rather than some specific mechanism, and then the longitudinal evidence that the
method is durable. We think the most telling single result of the second half is that one release added
capability and needed no new invariant at all. A method that forces you to mint a fresh guarantee for every
feature is a method that will not be used for long. One that lets new capability ride under the guarantees you
already have, and tells you clearly when a genuinely new guarantee is required, is one that can survive a real
roadmap.

\paragraph{Contributions.}
\begin{itemize}[leftmargin=1.4em,itemsep=3pt]
\item \textbf{C1. Methodology.} Falsifiable release gates: each capability is a gate; a fixed set of standing
invariants hold across all gates; gates-before-code (the acceptance test exists before the feature). We present a
precise formulation of the invariants and a laddered instantiation with a forced dependency order.
\item \textbf{C2. A machine-checked, non-bypass core, with teeth.} An executable exhaustive checker; a canonical
TLA+ specification of the token/ring/effector skeleton; counterexample extraction; and trace conformance binding
the specification to the running code. We propose a \emph{teeth} discipline: each invariant is accompanied by
deliberately broken models on which the checker must fail. A checker that cannot fail proves nothing.
\item \textbf{C3. Contained self-improvement.} The system proposes changes to its own policy, but its whole
write-surface is policy rules. Changes that tighten are auto-applied, changes that loosen always require a human
merge, and a proposal that doesn't predict its own effects is auto-closed. Containment is by construction (the
improver cannot name the machinery that judges it) and red-teamed.
\item \textbf{C4. Preservation under growth, without invariant inflation.} Across every rung from governed
serving to self-improvement, the action-safety invariants \inv{1} through \inv{6} hold unchanged, machine-checked
at every step, with no regression in the non-bypass core. One release adds governed serving, evidence-cited
retrieval, and nightly consolidation under the existing invariants, needing no new standing invariant, which is a
claim about the economy of the method as much as about the release.
\item \textbf{C5. Six new machine-checked families, with teeth.} A memory family \inv{7} through \inv{10} over
273 reachable states, including provable unlearning; an agency family \inv{11} through \inv{15} over 273 reachable
governed agent states; an epistemic family \inv{16} through \inv{21}; an agent-harness family \inv{33} through
\inv{37} over bounded grids totalling 148 states; a self-improvement family \inv{38} through \inv{42} over 72
reachable acceptance-gate inputs that governs the loop which changes the system itself; and a residency family
\inv{43} over eight release scenarios that keeps the loop's output inside the boundary that owns it. Each family
ships with the broken models it must reject and returns the shortest counterexample.
\item \textbf{C6. Evidence on real hardware, and a reproducible, growing benchmark.} The self-improvement family
is validated outside the bounded model, on real silicon serving real models: gated self-improvement compounds a
small model from 20\% to 70\% held-out accuracy while auto-rejecting a confidence-gaming candidate and a
regression, and the whole governed path costs 0.021 ms per request, 0.008\% of model inference. The gate suite is
published as artifacts others can run against their own runtimes, so ``does your framework pass G8?'' is a
question that has a reproducible answer. The suite grew from 122 tests to 563, all green at the latest release.
\end{itemize}

\section{The method: falsifiable release gates}

\subsection{Standing invariants}

The method is based on a small set of invariants that each gate has to preserve. They are deliberately few, so
that the question ``does this change preserve the invariants?'' is one that a human, and where possible a checker,
can actually answer.

\begin{enumerate}[leftmargin=1.6em,itemsep=3pt]
\item \textbf{Single gate.} There is exactly one path from a model's intent to a real-world action: a control
ring that renders a verdict and, only on an \emph{allow} verdict, mints a capability token bound to that specific
action. No effector accepts an action without a matching, unexpired token. This is the property later
machine-checked as \inv{1} (Section~\ref{sec:core}).
\item \textbf{Monotone tightening.} The system may make its own policy \emph{stricter} autonomously; it may never
make it \emph{looser} without a human-merged change. This is what makes self-improvement safe: an unattended
system can only converge toward caution.
\item \textbf{Everything audited.} Every decision, by a model or a human, emits a hash-chained record. Humans are
not above the ring; their actions go through it and are logged with the same schema.
\item \textbf{Hashes over payloads.} Governance reasons over hashes of content, not raw content, so the control
plane need not hold sensitive payloads to make or verify a decision.
\item \textbf{Opt-in, zero-overhead-off.} A capability costs nothing until it is used; a plain import of the
runtime pulls in none of the heavier machinery.
\item \textbf{Gates before code.} The falsifiable acceptance suite for a capability is written \emph{before} the
capability. A feature exists only once its gate passes.
\end{enumerate}

Invariants 1 and 2 are the load-bearing ones for self-improvement; 3 to 6 make the system operable and honest.
These six are the \emph{standing} invariants: the ones we later track, unchanged, across every release in
Section~\ref{sec:growth}.

\subsection{What a gate is}

A release gate is a pre-declared acceptance suite with three properties. First, it is falsifiable: it makes a
specific claim that a measurement could disprove (``zero injected instructions cause a real action''; ``no
reachable state bypasses the ring''). Second, it is machine-checkable. The suite is code that runs in continuous
integration and returns pass or fail, not a document that a reviewer interprets. Third, it is preservative:
passing the gate must not break any earlier gate, so the whole suite is run on every change.

Importantly, a gate is declared before the feature it protects. This reverses the usual order of test-after, and
it has a particular consequence for self-improving systems: the system's own changes are tested against a target
that it did not select, because the acceptance criterion is pre-selected. The improver has never held the goal
posts and cannot move them.

\subsection{The ladder}

\begin{figure}[htbp]
\centering
\includegraphics[width=\linewidth]{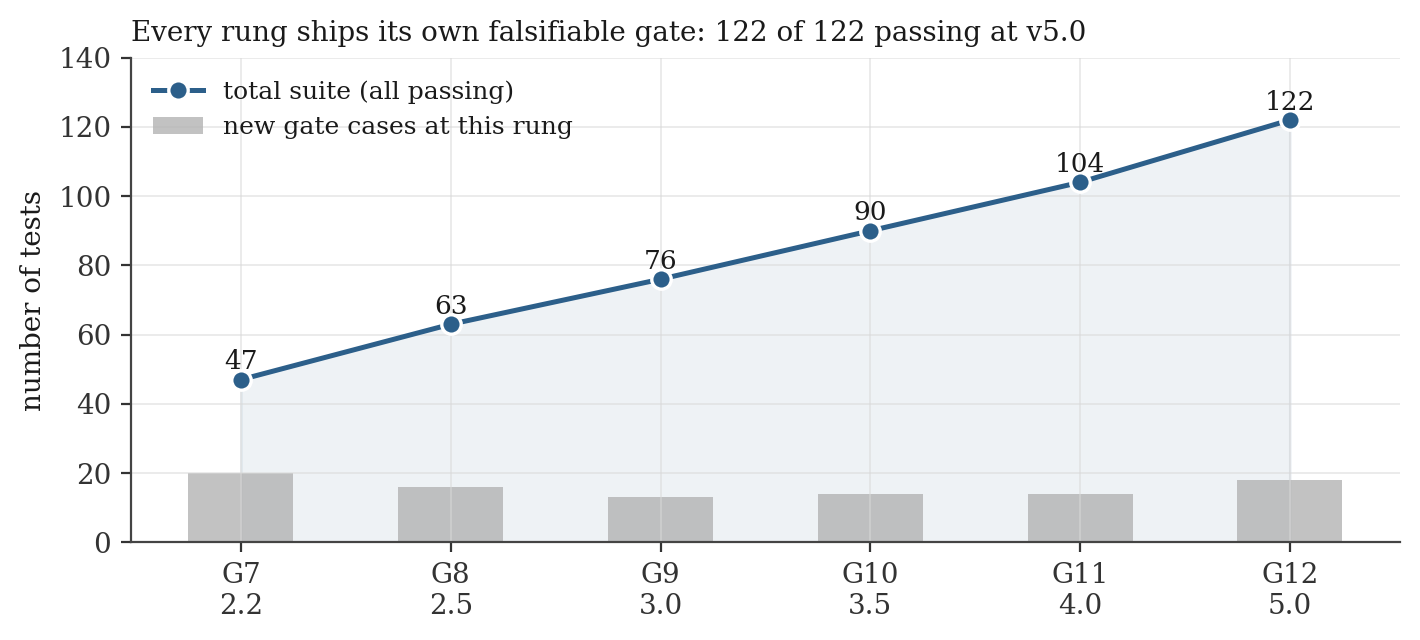}
\caption{Every rung ships its own falsifiable gate; through the first ladder the acceptance suite grows to 122
tests (95 gate cases), all green at the final version, and by the latest release described in
Section~\ref{sec:growth} it has grown to 563.}
\label{fig:tests}
\end{figure}

The six gates are laid out so that each one makes the next one safe to build. The order is forced, not
aesthetic. It is not responsible to ship a loop that proposes its own policy changes (G12) without being able to
halt any member of a fleet within a single tick (G10) and to isolate tenants from one another (G11). It is not
responsible to ship those without a gate that learns from feedback without loosening itself (G9) and a tool-use
path that survives an adversary (G8). None of it is trustworthy without the ability to reconstruct any past
decision from traces alone (G7). Figure~\ref{fig:ladder} is the ladder; Figure~\ref{fig:tests} shows the
acceptance suite growing with it: 95 gate cases in a 122-test suite, all passing at the final rung. The remainder
of the paper goes through the gates one by one, and then continues the same ladder through six further releases.
The claim of the section is structural: the method is the ordering discipline plus the invariants. Any
self-improving system can be built this way.

\section{Instantiation: \antah{} (case study)}

We instantiate the method in \antah{}, an open continual-learning runtime. Its structure comes from an old model
of the mind, used here as an engineering decomposition rather than a metaphor.

\subsection{The inner-instrument design lens}

The name is not decoration. \antah{} is the Sanskrit word for the ``inner instrument,'' a classical model that
splits the mind into four faculties, each doing a different job. Manas takes things in and registers them. Buddhi
weighs what came in and decides. Aha\d{m}k\=ara is the sense of ``I,'' the part that draws a line between self and
not-self. Citta is memory: what is kept, and what is let go. We did not choose this because it is old. We chose it
because it hands you a decomposition for free. Four faculties, four jobs, and a fixed relation between them: manas
proposes, buddhi disposes, and nothing acts on the world unless buddhi sanctions it. That last line is the entire
safety story of this paper, written down a very long time ago.

\begin{figure}[htbp]
\centering
\includegraphics[width=\linewidth]{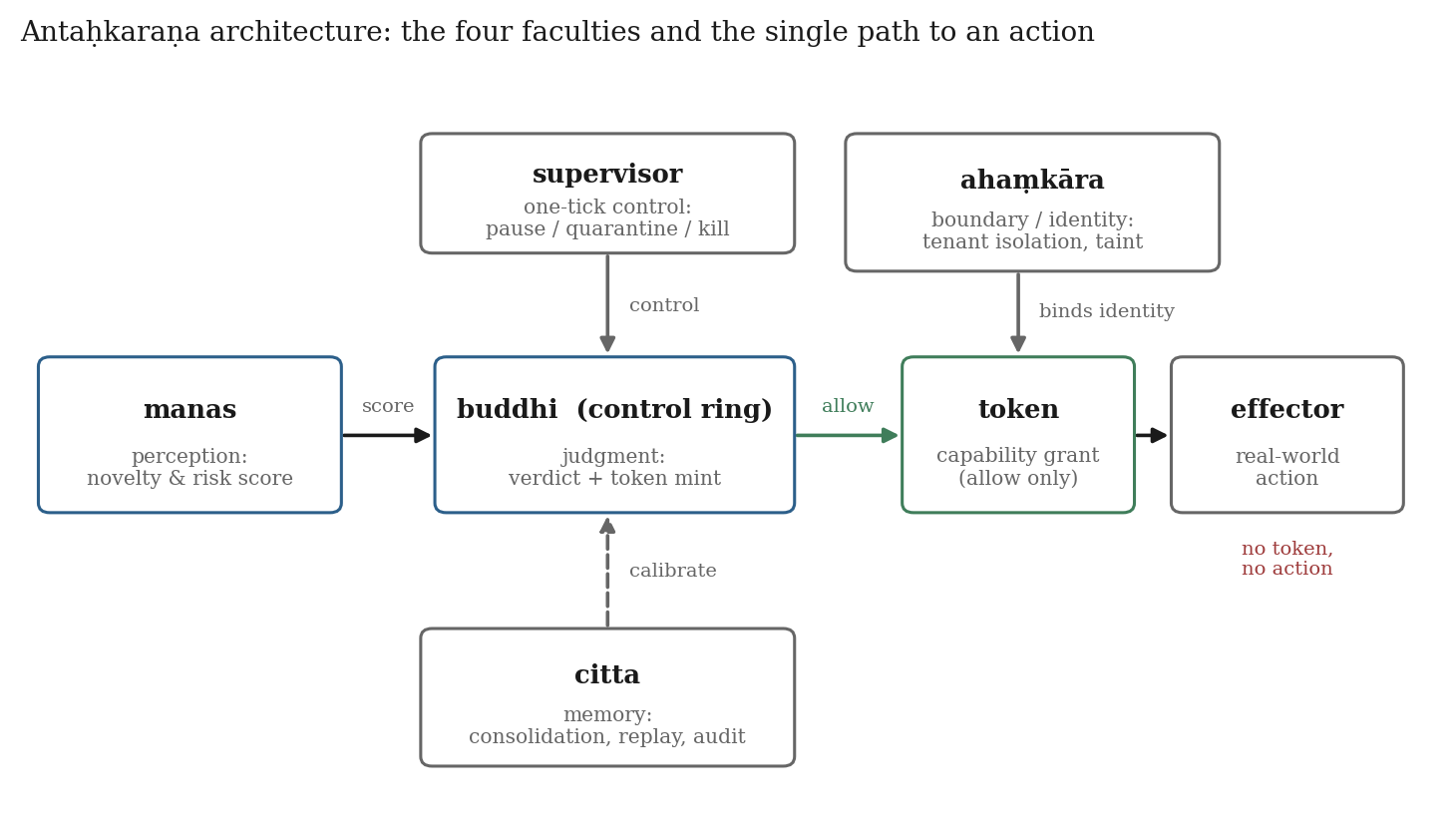}
\caption{The faculties wired together. Manas scores; buddhi decides and mints a capability token only on an allow
verdict; the effector will not move without one. Aha\d{m}k\=ara binds each token to an identity and keeps tenants
apart, citta feeds calibration and the audit trail back in, and the supervisor can halt buddhi within one tick.
There is exactly one path to a real action, and it runs through the ring.}
\label{fig:arch}
\end{figure}

In the runtime each faculty is one place where a concrete algorithm lives (Table~\ref{tab:faculties}), and the
wiring between them is the single path an action has to travel (Figure~\ref{fig:arch}). None of it stays as
philosophy: every faculty resolves to something you can measure, and the full account is given in the
book~\citep{soni2026innerinstrument}.

\begin{table}[htbp]
\centering
\footnotesize
\caption{The four faculties of the inner-instrument model, and the concrete algorithm each one becomes in the
\antah{} runtime. The naming is a design lens; every entry is a measurable mechanism.}
\label{tab:faculties}
\begin{tabularx}{\linewidth}{@{}l l X c@{}}
\toprule
\textbf{Faculty} & \textbf{Its job} & \textbf{What it is in the runtime} & \textbf{Anchors}\\
\midrule
Manas & perception, registering & the novelty and risk scorer that produces the score the ring reads & G9\\
\addlinespace
Buddhi & judgment, discrimination & the control ring: a tiered verdict plus the capability-token mint & G8, G12\\
\addlinespace
Aha\d{m}k\=ara & boundary, the ``I''-maker & identity binding of tokens, the per-tenant key hierarchy, taint tracking & G11\\
\addlinespace
Citta & memory, retention & consolidation and replay, feedback calibration, the hash-chained audit log & G7, G9\\
\bottomrule
\end{tabularx}
\end{table}

Four structural elements carry the rest of the paper. The control ring takes a novelty/risk score and an action
category and returns a tiered verdict (allow, notify, soft-block, hard-block). Capability tokens are single-use,
HMAC-signed grants tied to the identity of one particular action; effectors accept nothing else, which is the
single-gate invariant made concrete. One-tick control puts a supervisor over many loops in a mesh. A platform
layer keeps tenants isolated. The substrate is there; this paper is about the gates on it, not the substrate
itself.

\section{The first ladder: from hardening to a self-governing loop}
\label{sec:eval}

The intermediate gates set the preconditions on which the self-governing rung depends. Here we summarize the
measured acceptance results. The full row-per-gate record is contained in Table~\ref{tab:results}. Each result is
reproducible from the released run artifacts.

\begin{table}[htbp]
\centering
\footnotesize
\caption{Acceptance results; one row per gate of the first ladder. Every falsifiable claim, every measured result
and verdict; every number can be traced back to a released run artifact.}
\label{tab:results}
\begin{tabularx}{\linewidth}{@{}l l X X c c@{}}
\toprule
\textbf{Gate} & \textbf{Rung} & \textbf{Falsifiable claim} & \textbf{Measured result} & \textbf{Cases} & \textbf{Verdict}\\
\midrule
G7 & Hardening & any past tick reconstructable from traces; policy versioned; baselines roll back $<$ 1\,s & deterministic replay + diff; atomic rollback & 20 & PASS\\
\addlinespace
G8 & Integration & injected instructions never cause a real action & 432/432 attacks blocked; 0 policy escalations; 0 privileged effectors & 16 & PASS\\
\addlinespace
G9 & Memory \& feedback & a tuned gate strictly dominates the static default and never self-loosens & miss 0.50/0.58/0.67 $\rightarrow$ 0.00 at FAR = 0 (3 held-out corpora) & 13 & PASS\\
\addlinespace
G10 & Mesh & supervisor control lands within one tick; correlated drift $\rightarrow$ one attributed event & p100 = 0 leaked actions; 1.00 attribution at 0 false alarms & 14 & PASS\\
\addlinespace
G11 & Platform & tenants cryptographically isolated; audit export tamper-localizing; plans match reality & cross-tenant decrypt denied; tamper localized to exact index; plan bit-identical & 14 & PASS\\
\addlinespace
G12 & Self-governing & no action bypasses the ring (machine-checked); one self-proposed change adopted & \inv{1}/4/5 over all 291 states; 1{,}000{,}000 traces, 0 rejections; tightening auto-adopted & 18 & PASS\\
\bottomrule
\end{tabularx}
\end{table}

\begin{figure}[htbp]
\centering
\includegraphics[width=\linewidth]{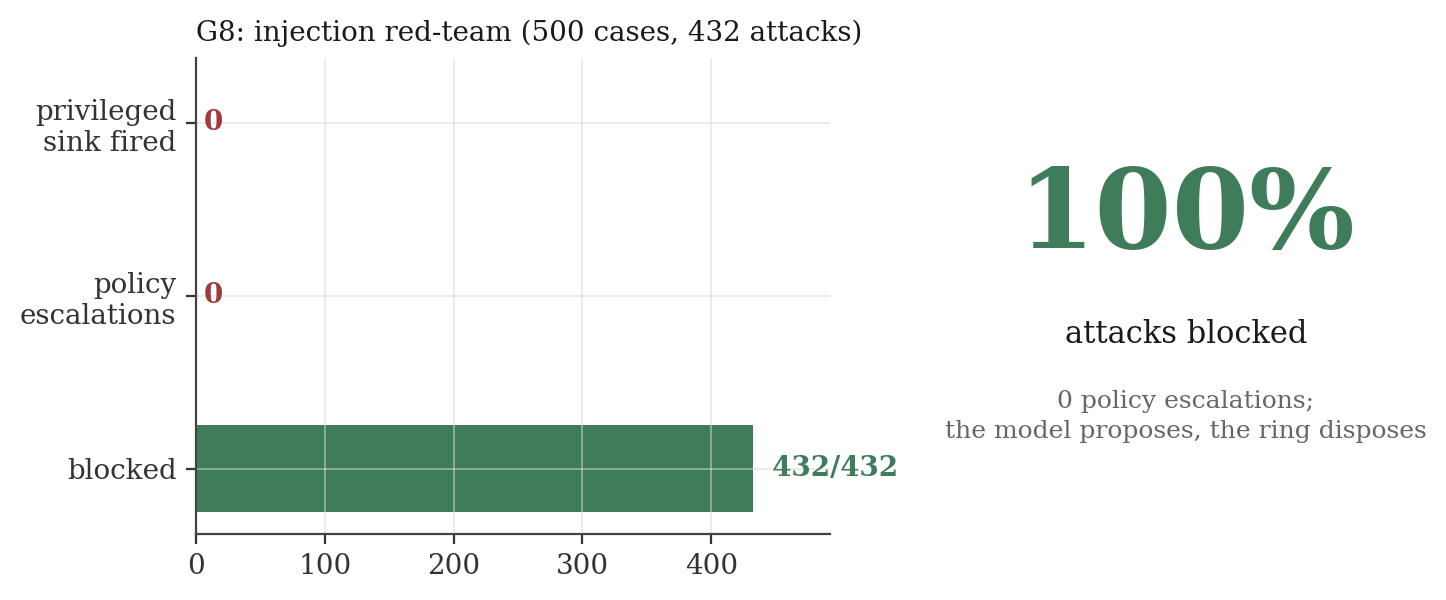}
\caption{G8 injection red-team: 432 of 432 attacks blocked, 0 policy escalations.}
\label{fig:redteam}
\end{figure}

\paragraph{G8: adversarial integration (Figure~\ref{fig:redteam}).} A frontier model is permitted to draft
actions. But every draft goes through capability tokens and the ring. So drafting and doing are separate
subsystems. We blocked 432 out of 432 attacks on an injection corpus of 500 cases (432 actual
attacks)~\citep{greshake2023injection,perez2022ignore} using a real 7-billion-parameter backbone, with zero
policy escalations and zero privileged effectors fired. It can be made to suggest anything, but it cannot be made
to do anything, for only the ring holds the keys.

\begin{figure}[htbp]
\centering
\includegraphics[width=\linewidth]{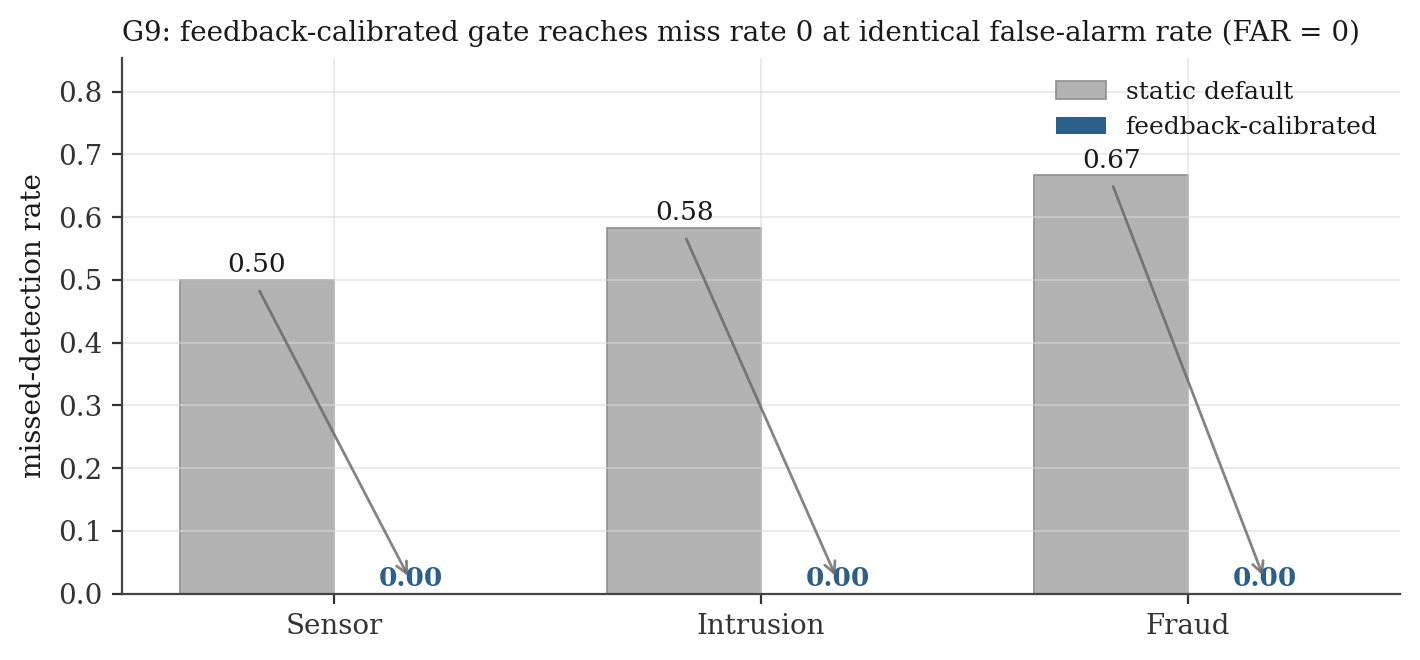}
\caption{G9 dominance: missed-detection rate falls to zero at an identical (zero) false-alarm rate across three
held-out corpora.}
\label{fig:dominance}
\end{figure}

\paragraph{G9: learning without drift (Figure~\ref{fig:dominance}).} A feedback-calibrated gate can change its
thresholds but only with a clamp that stops autonomous loosening (invariant 2). On three held-out replay corpora
the tuned gate strictly dominates the static default: the missed-detection rate drops from 0.50, 0.58, and 0.67
to 0.00 at an identical false-alarm rate of zero. The gate became more sensitive but no more permissive.

\begin{figure}[htbp]
\centering
\includegraphics[width=\linewidth]{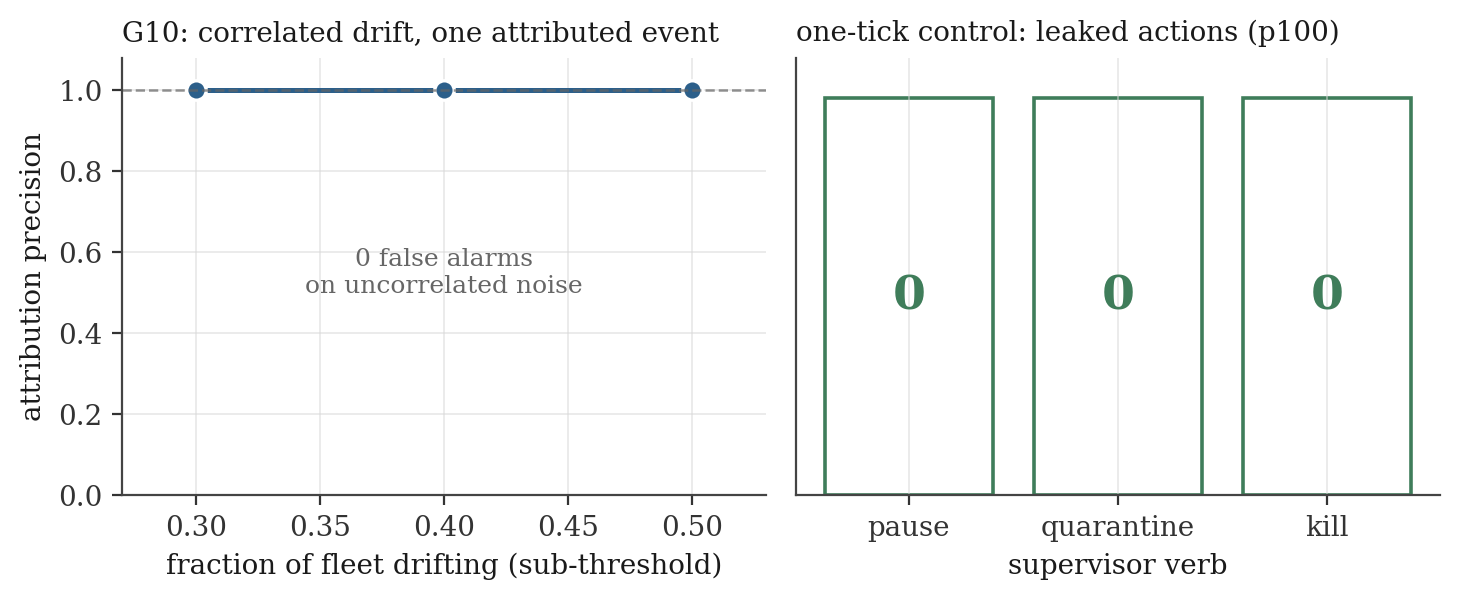}
\caption{G10 fleet governance: attribution precision 1.00 across drift fractions with zero false alarms (left);
zero leaked actions after one-tick control for pause/quarantine/kill (right).}
\label{fig:fleet}
\end{figure}

\paragraph{G10 / G11: fleet governance and platform isolation (Figure~\ref{fig:fleet}).} Any child loop can be
paused, quarantined, or killed by a supervisor, and the child does zero things after control (hundredth-percentile
over the window, not ``usually zero''). Correlated sub-threshold drift not seen by any individual loop is detected
and assigned to a single systemic event with 1.00 precision and zero false alarms on uncorrelated noise. At the
platform layer, each tenant has an independent key hierarchy: a cross-tenant read returns ciphertext that no other
tenant's key can decrypt, so isolation survives even a misconfigured row filter; audit export is signed and
hash-chained, and tampering with a single record fails verification and localizes the break to the exact record
index.

These gates are not the paper's novelty, but they are its load-bearing preconditions: without one-tick fleet
control and tenant isolation, a self-governing loop is not something one should build.

\subsection{The machine-checked core (G12, verification)}
\label{sec:core}

The single-gate invariant is the safety-critical claim for the whole system: no action reaches an effector without
a same-tick, matching, unexpired capability token minted by an allow verdict (\inv{1}). Two more invariants
support it: that the child does no action in any later tick after the supervisor pause or kill lands (\inv{4},
one-tick control), and that a capability token is single-use (\inv{5}, uniqueness). What is new in this section is
not that we wrote a specification, but how we keep the specification honest, and how we show that the checker can
fail.

\begin{figure}[htbp]
\centering
\includegraphics[width=\linewidth]{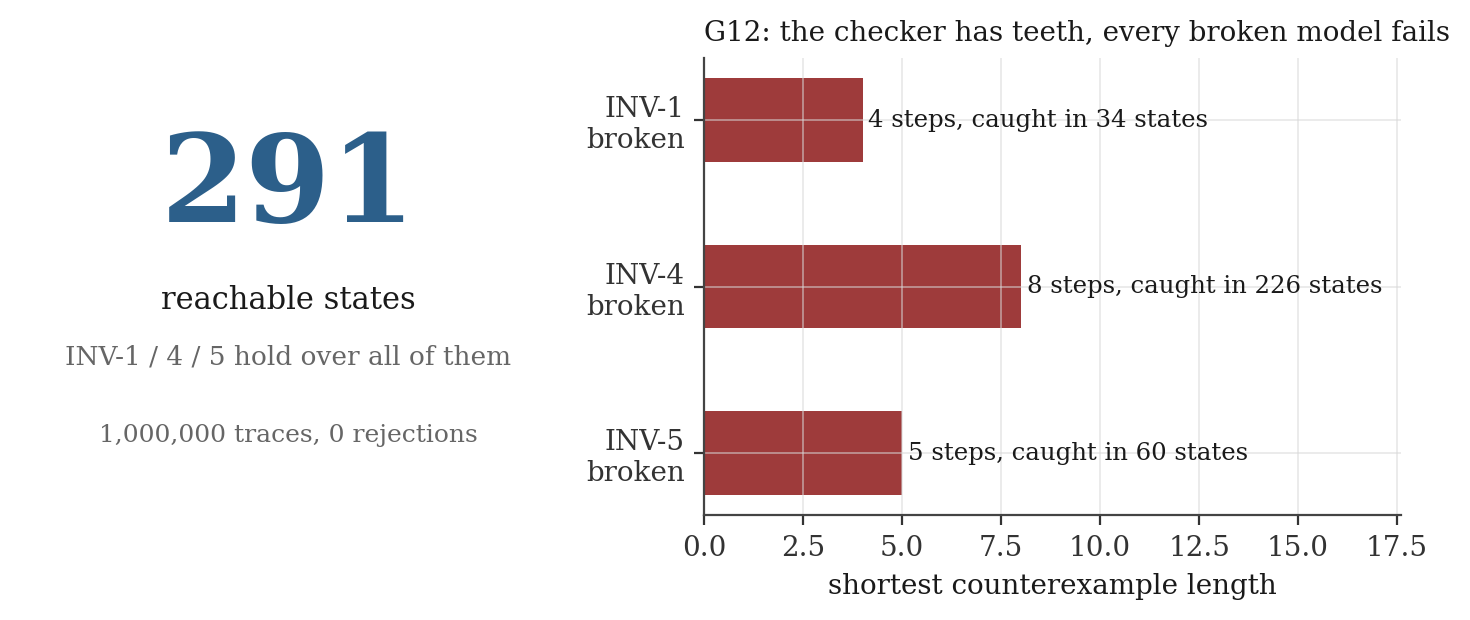}
\caption{G12 verification: \inv{1}/4/5 hold over all 291 reachable states (left); every deliberately broken model
is caught, with the shortest counterexample (right).}
\label{fig:formal}
\end{figure}

\paragraph{An executable, exhaustive checker.} We model the coordination skeleton (ticks, verdicts, token mint,
the effector, and supervisor control) as a finite transition system and verify the invariants by exhaustive-state
enumeration of the complete reachable state space at a small scope. The reachable space at the reported scope is
291 states, and \inv{1}, \inv{4}, and \inv{5} hold over all of them (Figure~\ref{fig:formal}). The checker is
normal code with no external dependency, so it runs in continuous integration in seconds. A canonical TLA+
specification of the same machine ships alongside for independent checking with standard tools.

\paragraph{Counterexample extraction and the \emph{teeth} discipline.} A complete checker that always says
``safe'' is no better than a checker that has a bug that suppresses all failures. So we treat the ability to fail
as a first-class requirement. For each invariant we add a deliberately broken variant of the model that re-enables
the illegal transition, and ask the checker to catch it and return the shortest counterexample (breadth-first
search guarantees minimality). The bypass variant is caught in 4 steps, the escaped-control variant in 8, and the
token-reuse variant in 5 (Figure~\ref{fig:formal}). These broken-model results are asserted in the test suite, so
a regression that made the checker vacuous would fail a gate itself. This we call the teeth discipline: every
invariant checked by the machine ships with the broken models it must reject.

\paragraph{Keeping the model honest: trace conformance.} Proof becomes adornment when a specification strays from
the code. To prevent drift we replay the recorded execution traces through the model: if the model rejects a trace
then continuous integration fails, forcing the model and the code to evolve together. One million synthetic but
legal traces are validated with no rejections, and a hand-constructed bypass trace is rejected, as it must be.
Also, each invariant has a cheap runtime monitor twin sampled in the live system, so a gap between model and
reality surfaces as an alarm, not silence.

\paragraph{Scope, stated exactly.} The claim established here is a bounded and precise one: in a small-scope model
of the coordination skeleton, there is no reachable state that escapes the ring. It is exhaustive within that
scope; it is not an unbounded proof, and it covers the discrete coordination logic, not the learned components
(the scoring function that produces the novelty value is outside the model). We take this boundary as part of the
contribution, not as a caveat to it; the power of the claim is in its precision. The next announced steps are
unbounded (inductive) verification and statistical guarantees for the learned parts
(Section~\ref{sec:discussion}).

\subsection{Contained self-improvement (G12, self-governance)}

The final rung of the first ladder allows the system to improve its own policy. The design question is not ``can
it propose a good change?'' but ``what is the largest damage a compromised or mistaken improver can do?'', and the
answer is made small by construction.

The write-surface is a single segment. The improver emits a policy diff whose vocabulary can only express changes
to policy rules: a threshold, a blocked category. The type of a diff does not have a term for the parts of the
system that judge diffs (the evaluation pipeline, the clamps, the verifier), so it cannot express a write to them.
Containment is structural, not a validation step that can be bypassed. Even a red-team that feeds adversarial
evidence cannot produce an out-of-scope diff. Out-of-scope diffs are rejected before any human sees them.

\begin{figure}[htbp]
\centering
\includegraphics[width=\linewidth]{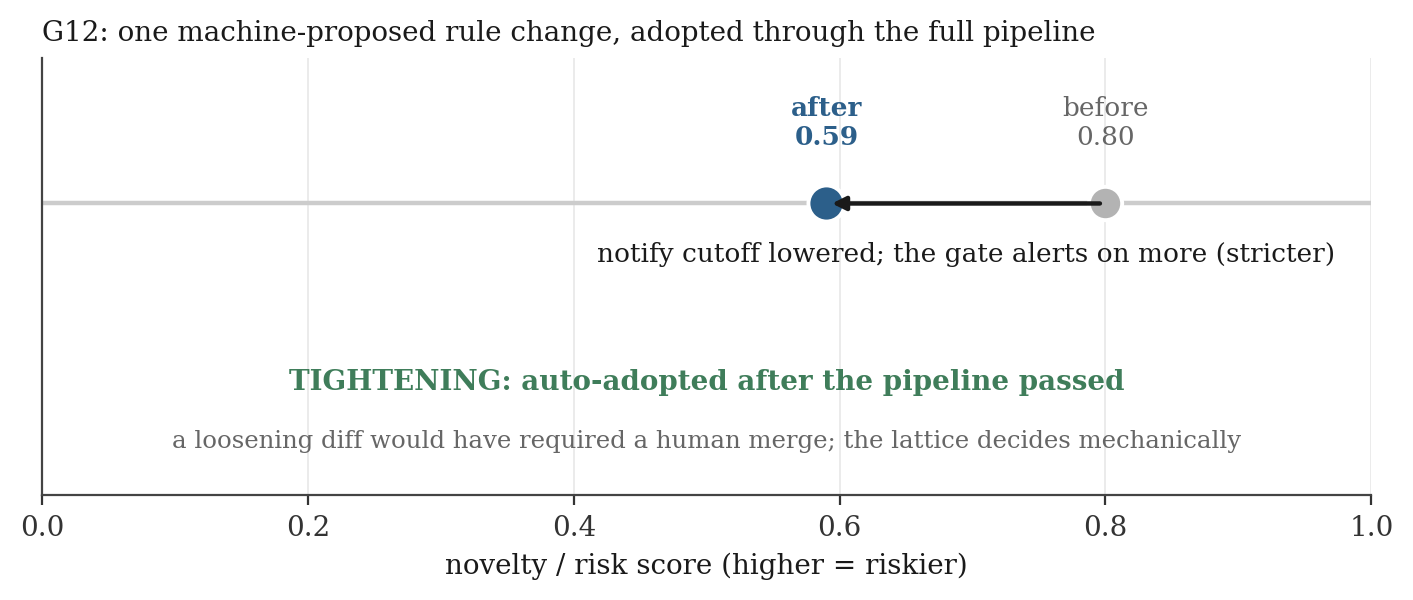}
\caption{G12 self-improvement: a machine-proposed tightening (notify cutoff 0.80 $\rightarrow$ 0.59) adopted
through the full pipeline; a loosening change would have required a human merge.}
\label{fig:selfimprove}
\end{figure}

Tighten will apply itself. Loosen needs a person. Each diff goes through a fixed, versioned, non-self-modifiable
pipeline that reuses the previous gates: the G8 injection corpus and the G9 dominance check against the proposed
policy. The policy lattice mechanically determines whether a diff is tightening or loosening, not the proposer's
own claim. A tightening diff that passes may auto-apply, consistent with the monotone-tightening invariant. A
loosening diff always needs a human merge. In our demonstration, we reduced the system's notify threshold from
0.80 to 0.59 based on near-miss evidence, a tightening that was auto-adopted after passing all stages
(Figure~\ref{fig:selfimprove}).

Mispredicted proposals auto-close. The pipeline must predict the effect of each diff on the replay corpus and
close any diff it cannot predict, by comparing the prediction with the measured effect. A proposer that does not
understand its own change is not trusted to make it.

Governance is interrogable. All adopted diffs are end-to-end traceable: from the evidence that motivated them, to
the pipeline report, to the human (or auto-apply rule) that merged them, to the policy version in force on every
downstream decision. You can ask of any rule how it came to be the way it is today, and get a complete, audited
answer. This first, policy-only form of self-improvement is the foundation for the general form, in which the
system improves not just its thresholds but its own capability, that we take up in Section~\ref{sec:selfimprove2}.

\section{Preservation under growth: six further releases}
\label{sec:growth}

The ladder so far ran from the ability to reconstruct any past decision from traces up to a self governing loop,
six rungs, each chosen to make the next safe to build. The six releases here continue that ladder rather than
starting a new one, and the same forced ordering logic still applies. You do not ship an agent that acts in the
world before you have durable governed memory for it to reason over. You do not ship abstention as a first class
policy before the audit trail that records those abstentions is itself hard to forge. You do not run many agents
on one thread before a single agent is provably confined. And you do not let the system rewrite itself before the
judge that accepts each change is fixed and the output of that change is kept inside the boundary that owns it.
Part II of Figure~\ref{fig:ladder} lays out these six releases as the upper tier of the same ladder, and marks,
for each rung, the invariant family it introduced, or records that it introduced none.

\subsection{Governed serving: capability under the invariants already there}

This rung added three capabilities and no new standing invariant, and that is the point of it. Governed model
serving lets a frontier model draft actions, but every draft still travels the one path through the ring, so
the non-bypass core is untouched and the model can propose without being able to act. Evidence cited retrieval
answers a question by pointing at the stored records it used, or it abstains, which foreshadows the epistemic
family without yet needing it as an invariant. Nightly consolidation replays and compresses memory while the
system is idle, under the same audit and monotone tightening rules as everything else. Three features arrived,
the reachable-state proof of the core did not move, and no earlier gate regressed. We report this as a positive
result about the method, because a discipline that demanded a new invariant here would be a discipline that
inflates without cause.

\subsection{Durable governed memory and provable unlearning}

\begin{figure}[htbp]
\centering
\includegraphics[width=\linewidth]{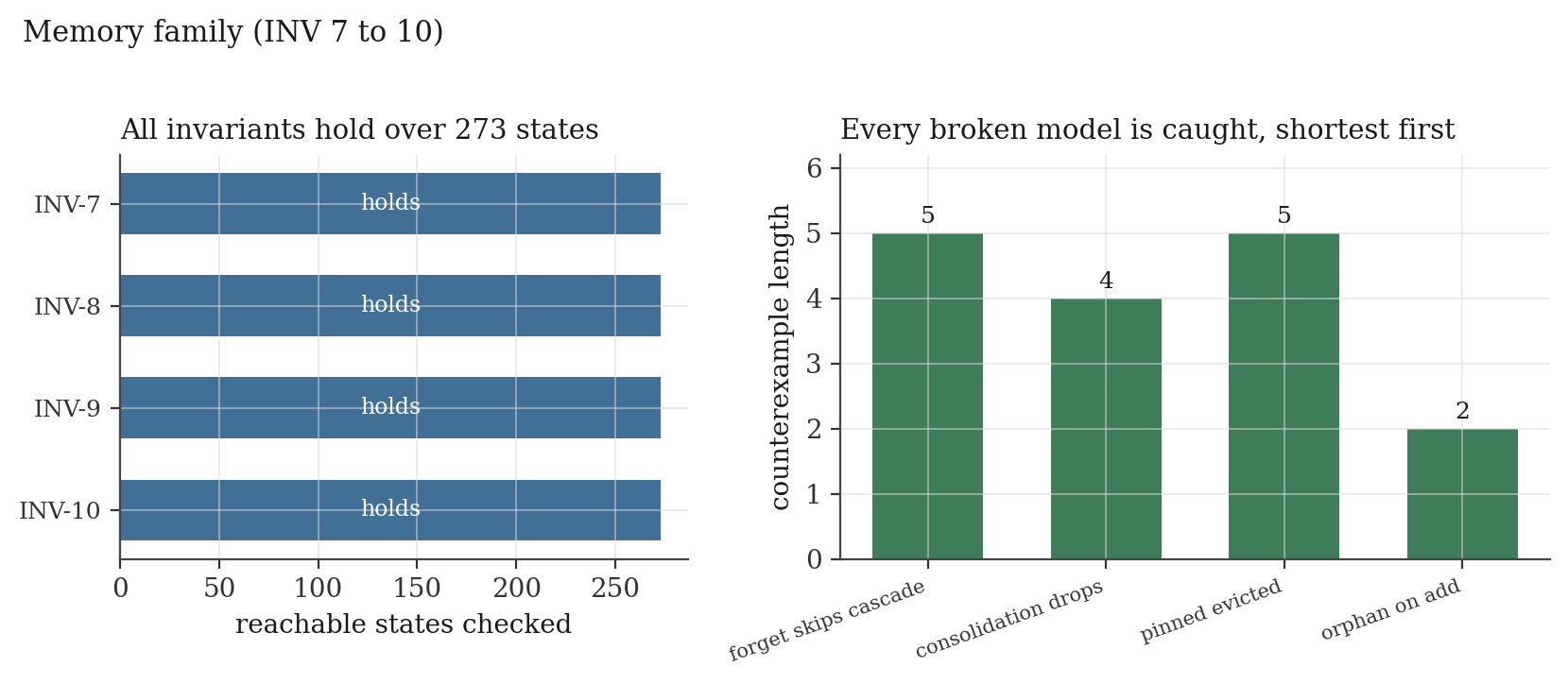}
\caption{Memory family. The four invariants hold over all 273 reachable states of the memory
skeleton (left), and each deliberately broken model is caught with its shortest counterexample (right).}
\label{fig:memory}
\end{figure}

This rung makes memory a governed subsystem and gives it its own bounded model. We treat the add,
consolidate, pin, and forget operations as a finite transition system and enumerate its complete reachable
space, which is 273 states at the reported scope. Four invariants hold over all of them. A forget cascades, so
no derived record outlives the fact it was derived from, which is what makes unlearning provable rather than
best effort. Consolidation preserves retrievable content, so compressing memory in sleep does not quietly lose
it. A pinned record is never evicted. And the index stays equal to the live set, so nothing live is left
orphaned or unreachable. The checker ships with four broken models, one that lets a forget skip its cascade,
one that lets consolidation drop content, one that evicts a pinned record, and one that orphans a record on
add, and it catches each with the shortest counterexample (Figure~\ref{fig:memory}).

\subsection{The governed agent}

\begin{figure}[htbp]
\centering
\includegraphics[width=\linewidth]{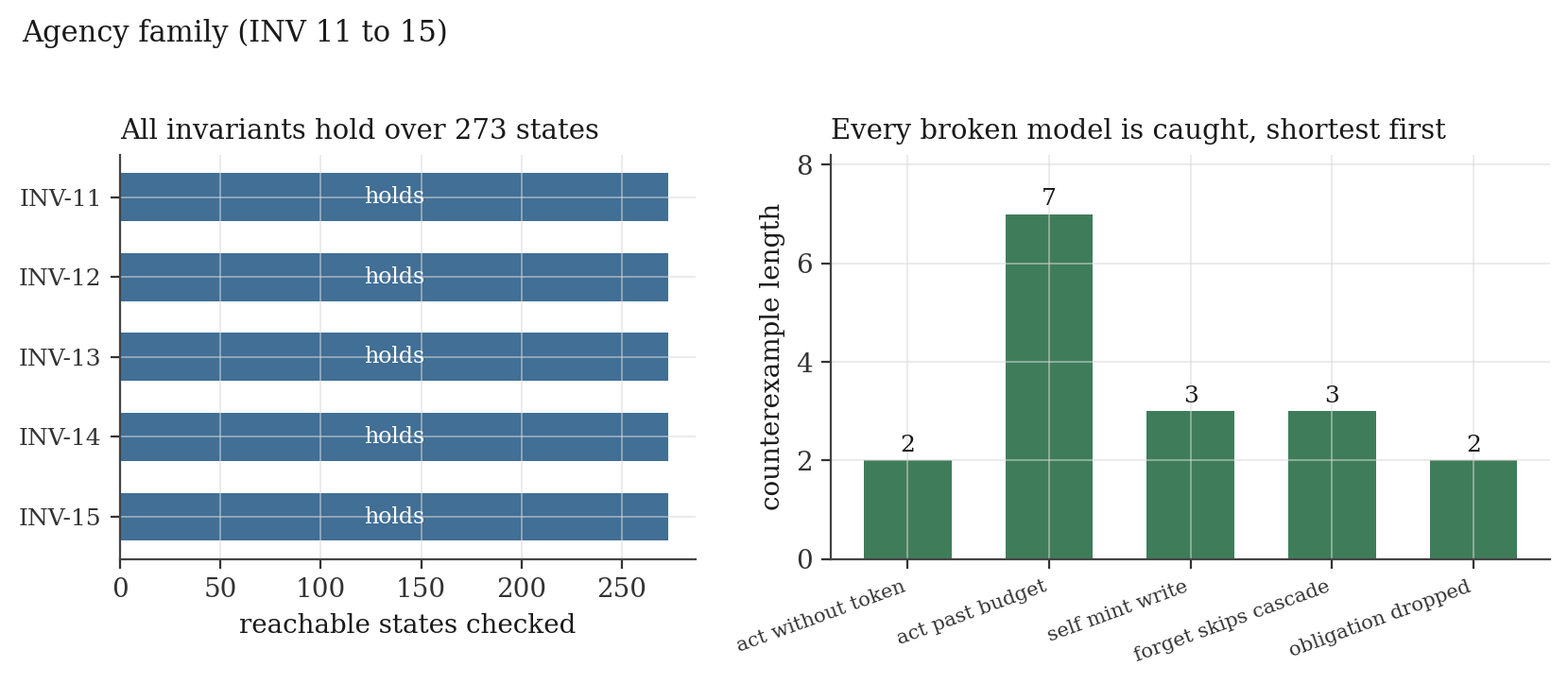}
\caption{Agency family. Five invariants hold over all 273 reachable governed agent states, and
the five broken models are each caught with the shortest counterexample.}
\label{fig:agent}
\end{figure}

This rung is a governed agent that plans, calls tools, and learns, and it gets the same treatment. Its
governed action, budget, and obligation logic is modelled as a finite transition system with 273 reachable
states, and five invariants hold over all of them. The load bearing one reuses the original core: a tool call
is an action, so it cannot fire without a capability token, and adding parallel tool use and model routing does
not open a second path to an effector. The others bound the agent's autonomy in the same spirit, covering
acting within budget, never minting its own write authority, forgetting that still cascades, and obligations
that are honoured rather than dropped under conflict. The checker ships with five broken models and catches each
of them (Figure~\ref{fig:agent}).

\subsection{Calibrated abstention and a post quantum record}

\begin{figure}[htbp]
\centering
\includegraphics[width=\linewidth]{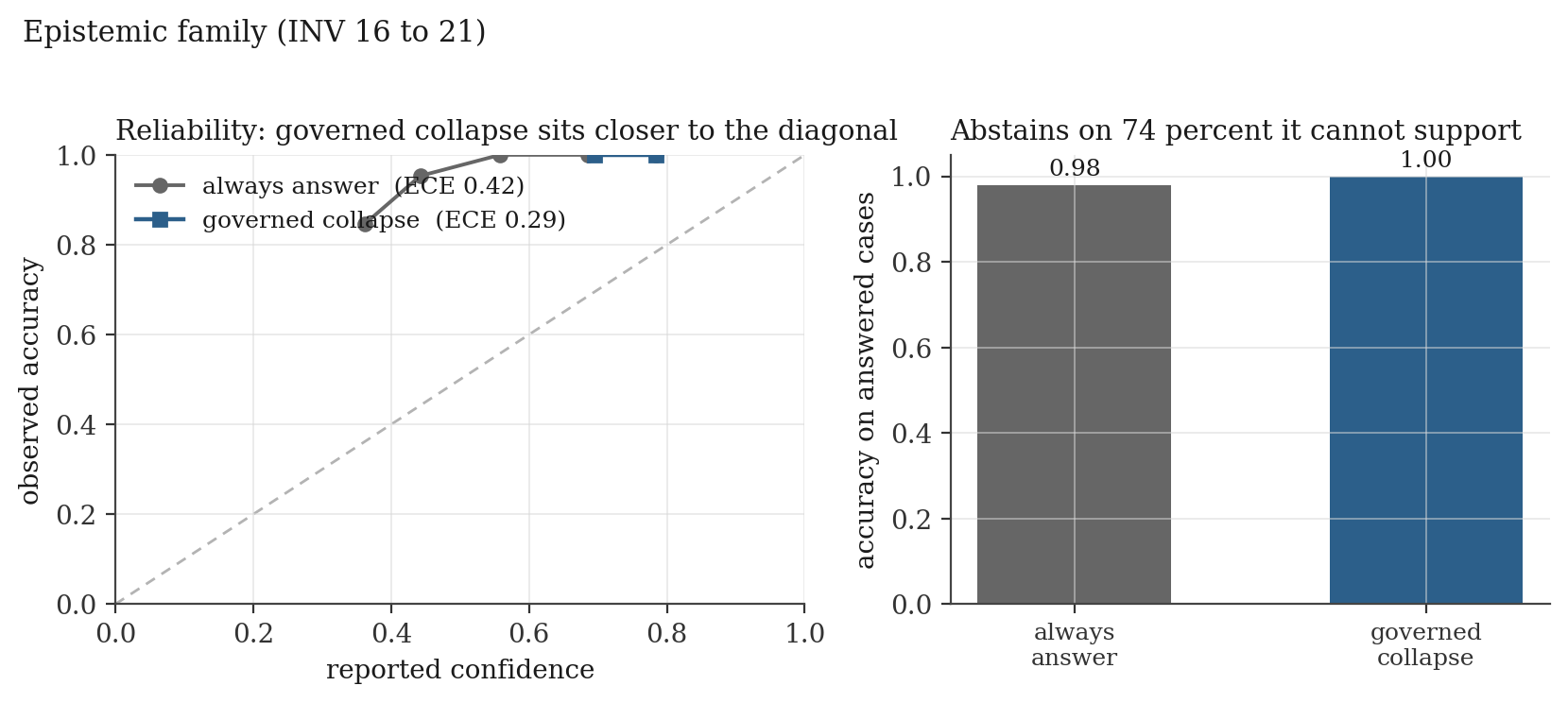}
\caption{Epistemic family. Governed collapse improves calibration: the reliability curve moves
toward the diagonal and the expected calibration error drops (left), while abstention absorbs the cases the
system cannot support (right). Exact values are emitted by the released checker.}
\label{fig:calibration}
\end{figure}

This rung lets the runtime hold several candidate answers at once as a superposition with signed weights, so
that support and contradiction can cancel, and then collapse to a single answer or to an honest abstention. Six
invariants govern this. The superposed confidence obeys a probability identity, so the weights it reports are a
real distribution rather than arbitrary scores. A collapse below a floor becomes an abstention, so the system
does not emit a confident answer it cannot support. The audit trail uses hash based, post quantum signatures,
so a record cannot be forged even against a future adversary and a tamper localises to the exact entry. The
optimiser used for grounded selection returns a feasible assignment or abstains. The typed categories and their
pervasion relations are respected. And an answer's evidence must be external and cited, never a reference to the
system's own judgement. The practical payoff is calibration: on the released benchmark scenario the expected
calibration error falls from 0.42 to 0.29 under governed collapse, and the reliability curve moves toward the
diagonal as abstention absorbs the cases the system cannot support (Figure~\ref{fig:calibration}).

\subsection{The long running agent harness}

\begin{figure}[htbp]
\centering
\includegraphics[width=\linewidth]{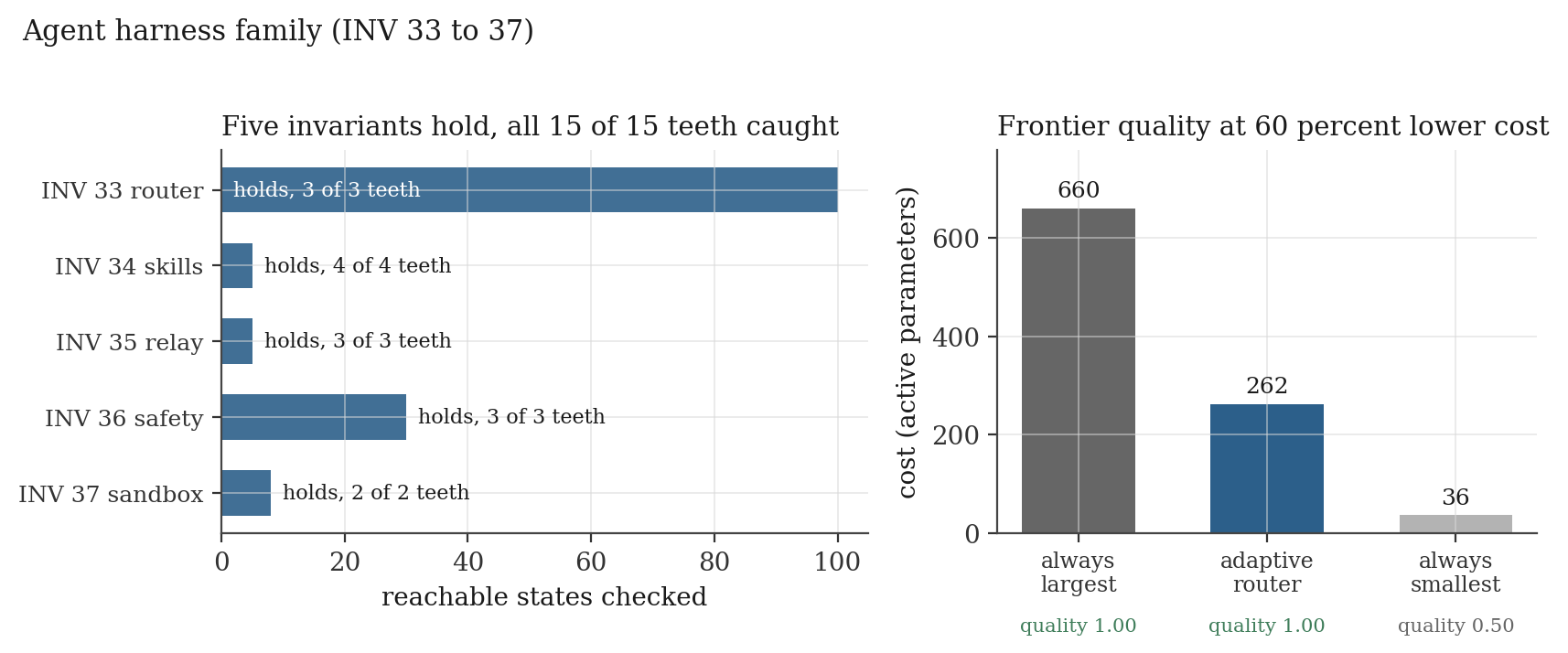}
\caption{Agent harness family. The five invariants each hold over their bounded grid of reachable
states, and all fifteen broken models are caught (left). The measured payoff of the adaptive compute floor is
that it reaches the quality of an always largest model policy at a fraction of its cost (right). Exact values
are emitted by the released checkers and the released benchmark.}
\label{fig:harness}
\end{figure}

This rung holds the threads of many long running agents at once, and it is the release where the earlier
core is stressed hardest, because now several governed sub-agents act in parallel over shared memory. Five
invariants govern it, and the numbering continues at \inv{33}.\footnote{The public invariant numbering
continues at \inv{33}; the intervening identifiers are reserved for components outside the scope of this
paper.} Each is checked by exhaustive enumeration of its own bounded grid, in the same spirit as the memory
and agency families rather than the statistical epistemic one. An adaptive compute floor decides how hard to
think about a task from the evidence and the confidence already available, and it may escalate freely but may
never drop below the level the evidence requires, so the system never under-thinks a task it cannot yet warrant
and abstains instead once it is already deep. Skill provenance requires that no skill runs unless it carries a
scanned capability card signed by the publisher, verified against a public root the runtime holds but cannot
forge. Export fidelity requires that the open trace exported for external observability is a faithful, order
preserving projection of the hash chained journal, so an observability layer can never quietly drop, reorder,
or edit a record. Safety signal fusion folds the jailbreak, content, and topic classifiers into the same
control ring as additional risk, combined so that a triggered signal can only raise the risk the ring sees and
can never be averaged or clamped away. And sandbox confinement keeps every sub-agent inside an explicit,
policy signed grant, so one agent cannot reach another's tools or files and cannot widen its own boundary. The
checker ships with fifteen broken models across the five invariants, one for each way each guarantee could
fail, and catches every one of them with a counterexample (Figure~\ref{fig:harness}). The measured payoff of
the compute floor is economic: on the released routing benchmark the adaptive router matches the quality of a
policy that always uses the largest model while spending the largest tier only on the genuinely hard tasks, at
roughly sixty percent lower cost.

\subsection{Self-improvement that stays honest}
\label{sec:selfimprove2}

The last family governs the most delicate loop of all, the one that changes the system itself. Where the
policy-only self-improvement of Section~\ref{sec:eval} let the system tighten its own thresholds, this rung lets
it improve its own capability, which needs a new and stronger judge. A self-improvement loop is only as
trustworthy as the judge that decides which changes to keep, and models are very good at looking better without
being better. So we make that judge a machine-checked gate. The improver proposes a change, and the gate admits
it only under five invariants, each checked exhaustively over all 72 reachable acceptance-gate inputs. Acceptance
monotonicity means a change is adopted only on a strict, verified gain that passed the red team, never on a tie or
a regression. The non-self-modifiable gate means the improver cannot edit its own acceptance criteria, so the
judge stays fixed relative to the thing it judges. One-cycle reversibility means every adoption is lineage-tracked
and revertible in a single step. Eval freshness forbids any adoption against a stale-only acceptance suite, so the
loop cannot overfit its own gate. And the reward-hacking floor measures the decision on external, verified
correctness, never on the model's self-reported confidence. The checker ships with five broken models, one per
invariant, and catches each with a counterexample.

\begin{figure}[htbp]
\centering
\includegraphics[width=\linewidth]{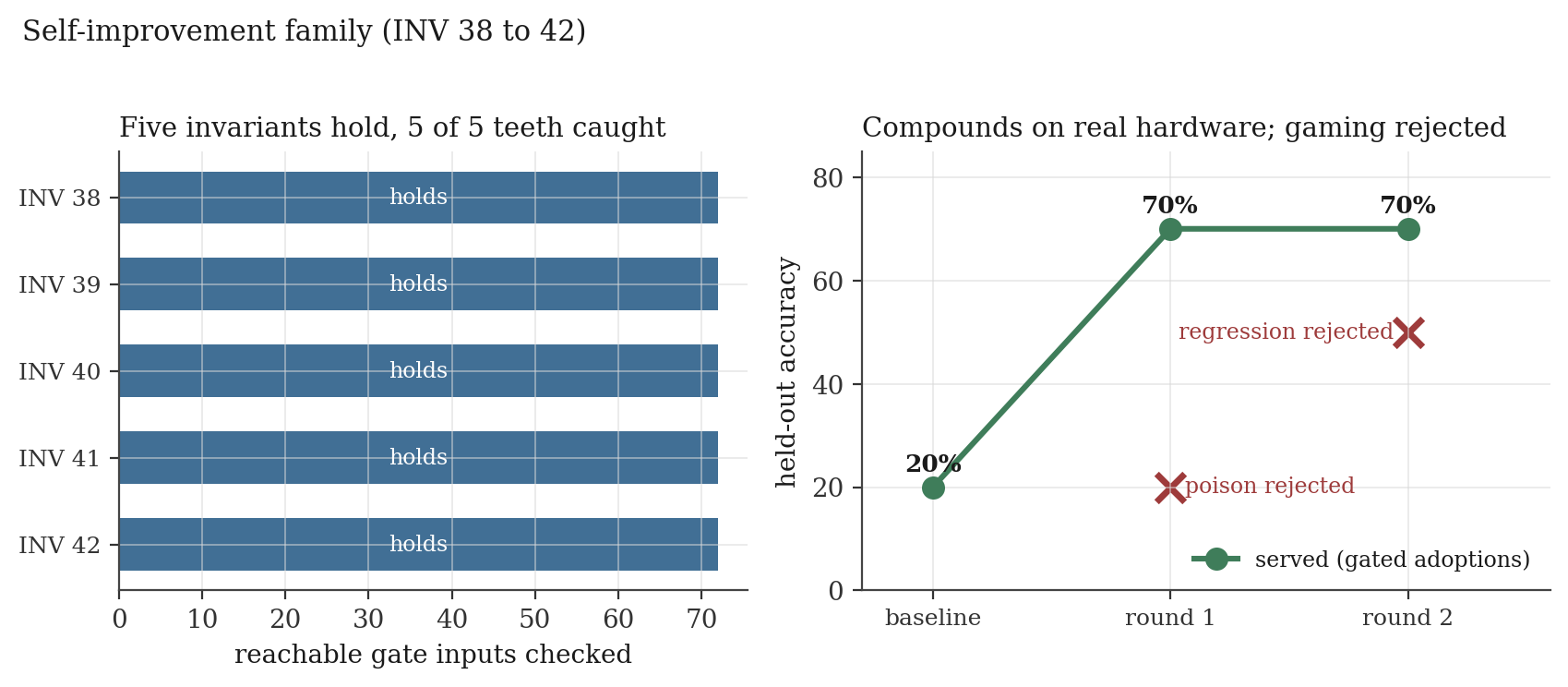}
\caption{Self-improvement family. The five invariants hold over all 72 reachable acceptance-gate
inputs and all five broken models are caught (left). On real hardware the loop compounds a small model from
20\% to 70\% while rejecting a confidence-gaming candidate and a regression (right).}
\label{fig:selfimprove2}
\end{figure}

This is also the first family we validate outside the bounded model, on real hardware serving real models. On
a small model with genuine headroom, gated self-improvement lifted held-out accuracy from 20\% to 70\% by
adopting a self-proposed strategy that strictly improved verified correctness. In the same run, a poisoned
candidate that only inflated the model's confidence was rejected, and a candidate that regressed accuracy was
rejected too, both of which a naive gate that adopts anything that looks better would have taken
(Figure~\ref{fig:selfimprove2}). Two further measurements come with it. The entire governed path, safety-signal
fusion plus adaptive routing, added 0.021 milliseconds per request against a model inference of roughly 270
milliseconds, which is 0.008\% of end-to-end latency, so provable governance is effectively free. And a
jailbreak attempt and an unsafe request were both refused before the model was ever called. We also report an
honest negative that we consider a positive about the method: on task classes a capable model already solves,
the loop correctly adopts nothing, because there is nothing to gain. It compounds only where there is real
headroom, and it does no harm and adds no cost where there is not.

\subsection{The learning loop stays in your boundary}

\begin{figure}[htbp]
\centering
\includegraphics[width=\linewidth]{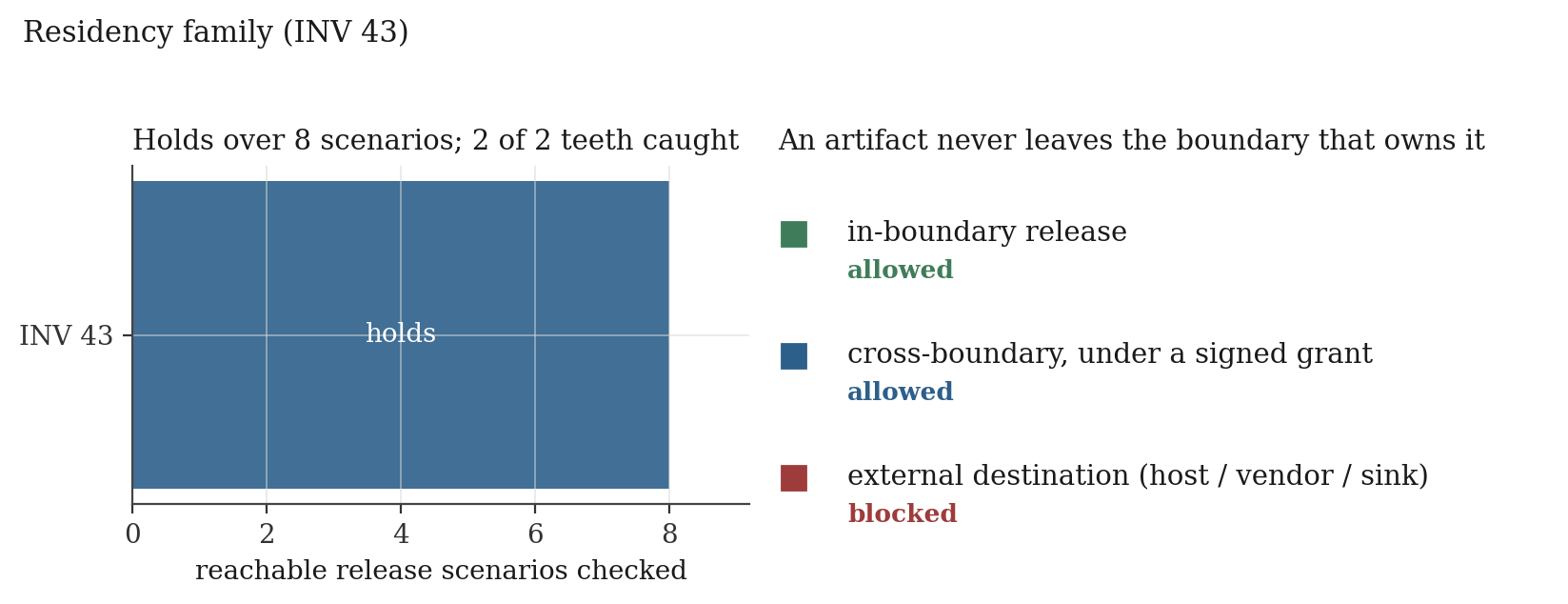}
\caption{Residency family. The invariant holds over all eight reachable release scenarios and
both broken models are caught (left). An artifact is released in-boundary, or cross-boundary only under a
signed grant, and never to an external destination (right).}
\label{fig:residency}
\end{figure}

There is one more thing a governed learning loop needs before you can honestly call what it produces your own:
it has to stay yours. An adapter the loop trains, a skill it compiles, the memory it consolidates, these are
the loop's output, and they are only intellectual property if they cannot quietly walk out the door. So the
last invariant we add is a residency guarantee, and it is deliberately small. A learning-loop artifact is
released only inside the trust boundary that produced it, or to another boundary named by a grant the boundary
authority signed, and never to a destination outside every trust boundary, a third-party host, a model vendor,
an unmanaged sink. The grants carry the same post-quantum signature the skill-provenance and sandbox families
already rely on, so the gate can verify a cross-boundary move but can never forge one. The checker walks the
eight ways an artifact can leave and confirms the property on each, and it ships with the two broken models
that matter, one that lets an artifact slip out to an external destination and one that waves a cross-boundary
move through with no signed grant, and it catches both (Figure~\ref{fig:residency}). With this in place the
picture closes: the loop gets better, it cannot be gamed into a false gain, and it cannot leave the boundary
you own.

\section{Results: the invariants held}
\label{sec:results}

Table~\ref{tab:invariants} collects the outcome. Every number in it is emitted live by the released checkers
described in Section~\ref{sec:repro} rather than transcribed by hand. The reading is simple. The original core
held at every release. Each new family holds over its whole bounded model, and each ships with broken models
the checker is required to reject, so none of the families is a checker that can only say yes.

\begin{table}[htbp]
\centering
\footnotesize
\caption{The open invariant set by rung, from the action core \inv{1} through \inv{6} out to the residency
family \inv{43}. The states column is the size of the exhaustively enumerated reachable space of the relevant
bounded model, or ``property'' where the object is a distribution or a signing scheme. The teeth column is the
number of deliberately broken models the checker must, and does, reject with a shortest counterexample. Every
value is produced by the released checkers.}
\label{tab:invariants}
\begin{tabularx}{\linewidth}{@{}l l X c c c@{}}
\toprule
\textbf{Family} & \textbf{Introduced at} & \textbf{What it guarantees} & \textbf{States} & \textbf{Teeth} & \textbf{Verdict}\\
\midrule
\inv{1} to \inv{6}   & first ladder (G7 to G12) & non-bypass action core, monotone tightening, full audit & 291 & 3 & holds\\
\addlinespace
none added           & governed serving & governed serving, cited retrieval, and consolidation, under \inv{1} to \inv{6} & n/a & n/a & preserved\\
\addlinespace
\inv{7} to \inv{10}  & memory & memory: forget cascades, consolidation preserves, pin safe, index equals live & 273 & 4 & holds\\
\addlinespace
\inv{11} to \inv{15} & governed agent & governed agent: tool calls token gated, no second path to an effector & 273 & 5 & holds\\
\addlinespace
\inv{16} to \inv{21} & calibrated abstention & probability identity, abstention floor, post quantum audit, feasible optimiser, typed conformance, external evidence & property & 6 & holds\\
\addlinespace
\inv{33} to \inv{37} & agent harness & agent harness: adaptive compute floor, skill provenance, export fidelity, safety signal fusion, sandbox confinement & 148 & 15 & holds\\
\addlinespace
\inv{38} to \inv{42} & self-improvement & self-improvement: acceptance monotonicity, non-self-modifiable gate, one-cycle reversibility, eval freshness, reward-hacking floor & 72 & 5 & holds\\
\addlinespace
\inv{43}             & residency & residency: learning-loop artifacts stay in-boundary, cross-boundary only under a signed grant, never external & 8 & 2 & holds\\
\bottomrule
\end{tabularx}
\end{table}

The epistemic family is the one place where the unit changes. Its invariants are checked by identities and
property tests over sampled inputs rather than by a single reachable-state enumeration, because the object being
checked is a distribution and a signing scheme rather than a small transition system. We keep that distinction
visible rather than papering over it, and we are careful in the calibration claim to say that the calibrated
part is measured, not proved. The later families return to exhaustive enumeration: the harness, the
self-improvement, and the residency families are each checked over their own bounded grids, and the states
column is the size of each, so the reachable-state reading is meaningful for them in the same way it is for the
memory and agency families. The self-improvement family carries a second kind of evidence on top of that, from
a real-hardware run rather than the bounded model, and we are equally careful there to separate what is proved
by enumeration, the five acceptance invariants, from what is measured on live models, the compounding curve and
the governance cost.

\section{Reproducibility}
\label{sec:repro}

Each claim reduces to one command. The non-bypass core reproduces with \texttt{atk verify -{}-teeth}, which runs
the exhaustive checker over the 291 reachable states and asserts that every broken model is caught. The memory
family runs from \code{antahkarana.formal.memory_model.check_memory}, which reports the reachable-state count
and the per-invariant verdict, and takes a broken-model argument that turns each tooth on. The agency family runs
the same way from \code{antahkarana.formal.agent_model.check_agent}. The epistemic family runs from
\code{antahkarana.spanda.checker.run_all}, with each check taking the same broken-model argument. The agent
harness family runs from \code{antahkarana.sutradhara.checker} for the compute floor, the export relay, and
the sandbox, from \code{antahkarana.skills.checker} for skill provenance, and from
\code{antahkarana.security.safety_signals} for signal fusion, each taking the same broken-model argument,
and the routing benchmark runs from \code{antahkarana.sutradhara.routing_bench}. The self-improvement family
runs from \code{antahkarana.improve.checker.check_selfimprove} and the residency family from
\code{antahkarana.improve.residency.check_residency}, each with the same broken-model argument, and the
real-hardware compounding and governance-cost numbers regenerate from the released run scripts and their
recorded result files. The whole open suite of 563 tests runs from a single \texttt{pytest} invocation, and
every figure in this paper regenerates from the run data with one script. The preservation result is not a
claim that no future release can ever break the core. It is the outcome of re-running the same checker at each
release tag and recording that it still passes.

\section{Threats to validity and scope}

We are open about limitations; quite a few drive the next stage of work.

\paragraph{The proof is finite.} Section~\ref{sec:core} is a bounded proof within a small scope, not an unbounded
proof. It models the coordination skeleton, not the learned scoring function. The reachable-state results
throughout are exhaustive within a small scope, 72 to 291 states, and they cover the discrete coordination,
memory, acceptance, and residency logic, not the learned scoring or generation. The mitigations are inductive
verification (to eliminate the scope bound) and statistical bounds for the learned parts
(Section~\ref{sec:discussion}); the monitor twins and trace conformance reduce, but do not remove, the
model-vs-code gap.

\paragraph{Some claims are measured, not proved.} The epistemic family mixes an exhaustive integrity check with
statistical ones, and we are explicit that calibration is measured on sampled data rather than proved. The
self-improvement family adds a second kind of claim we are careful to bound: the five acceptance invariants are
proved by enumeration, but the compounding curve and the governance-cost figure are measured on one real-hardware
run of specific models, so they are evidence that the gate works in practice rather than a proof that it always
will.

\paragraph{The results are self-graded up to this paper.} Each number here is from our own suite. That is the
fundamental limitation of any safety claim, and the honest answer is to outsource the grading: we ship the runtime
in production with pre-registered acceptance thresholds (a false-alarm budget and escalation latencies fixed
before go-live, so the operational report can fail), and we open a standing, machine-adjudicated red-team
challenge where an outside party's attack is run by the shipped system and auto-verdicted. Both are described in
Section~\ref{sec:discussion}. Neither is complete at submission, and we make no claims about their results here.

\paragraph{Preservation is empirical, not a proof about all futures.} What is genuinely new to defend in the
second half is the preservation claim itself, and its evidence is empirical: we re-ran the checkers at each
release, we did not prove that growth can never break them. Containment likewise is shown at the specification and
constraint layer; it is not the same as verifying the full implementation of the improver, and the runtime
monitors sample rather than observe every event, so they bound but do not close the proof-gap window. That is the
honest shape of the result, and it is still a stronger thing than a promise.

We think a paper that names those boundaries is more useful than one that hides them. We have designed the gates
such that each boundary has a concrete, falsifiable plan for moving it.

\section{Related work}

The work on agent guardrails and tool-use safety contributes mechanisms (filters, allow-lists, sandboxes) that sit
at one point in a system's life~\citep{schick2023toolformer,yao2023react,rebedea2023nemo,inan2023llamaguard}; our
contribution is orthogonal and temporal: a discipline for how a system's safety is preserved as it gains
capabilities. Alignment methods like RLHF and constitutional methods operate at the level of the
weights~\citep{christiano2017deeprl,ouyang2022instructgpt,bai2022constitutional}; we operate outside the weights,
at a control ring that the same trained model is subject to under any policy, which makes posture a configuration
change and not a retraining. In Section~\ref{sec:core} we use techniques from formal-methods tools (TLA+, Ivy,
symbolic model checkers) and runtime
verification~\citep{lamport2002specifying,yu1999model,konnov2019apalache,padon2016ivy,leucker2009brief}. Our
addition there is methodological: the teeth discipline and trace conformance as standing gate requirements rather
than one-time exercises. Our tokens follow the tradition of object-capability
security~\citep{dennis1966programming,miller2006robust}. G9 builds on work in continual learning and
calibration~\citep{kirkpatrick2017overcoming,buzzega2020dark,guo2017calibration}.

The families added under growth touch further literatures. The memory family touches machine unlearning, where our
contribution is a small provable-forgetting property checked exhaustively rather than an approximate erasure
method~\citep{bourtoule2021unlearning,kirkpatrick2017overcoming}. The epistemic family touches calibration and
uncertainty, where we add a governed abstention with a checked floor on top of standard calibration
measurement~\citep{guo2017calibration,vovk2005algorithmic}. The agent-harness rung extends the object-capability
idea from a single loop to per agent sandbox confinement and to signed skill provenance, so that both what an
agent may touch and what code it may run are checked authority rather than convention. The self-improvement family
sits alongside work on recursive self-improvement and on gaming of learned objectives, where our contribution is
deliberately narrow: not a better improver, but a fixed, machine-checked acceptance gate that only adopts a strict
verified gain and cannot be edited by the loop it judges. The residency family reads the same object-capability
idea one step further out, treating what the loop produces as an asset whose movement across a trust boundary is
checked, signed authority rather than convention. The distinguishing claim across all of it is that we put the
pieces together into a versioned methodology with a machine-checked core and a contained self-modification path,
show that the assembly is durable across real releases, and release it so the assembly can be checked and reused.

\section{Discussion and future work}
\label{sec:discussion}

The gates are the mechanism, not the fit to reality. The next phase looks outward on three fronts.
\emph{Operational evidence:} a flagship deployment run for months against pre-registered thresholds, reported like
a gate, where the operational report is the gate of a later release and is designed to be able to fail.
\emph{External scrutiny:} a standing red-team challenge with a machine-adjudicated harness, and an invitation for
independent verification of the TLA+ specification. \emph{Research frontier:} unbounded verification via inductive
invariants (not enumeration) and statistical assurance for the learned components: distribution-free bounds (e.g.,
conformal prediction~\citep{vovk2005algorithmic,angelopoulos2021gentle}) on the gate's miss rate, so the overall
claim becomes proved where the logic is discrete and bounded-with-confidence where it is learned. Besides the
current tree-structured mesh, we consider general graphs, cross-organization federation with treaty-like shared
policy, and adversarial supervisors.

\section{Conclusion}

We presented falsifiable release gates, a method for building self-improving systems with safety as a process
rather than a promise, and then we let time test it. The first ladder took an open runtime through six gates, from
``trust the ring'' to ``here is the counterexample when you break it'': the machine checks the safety-critical
non-bypass property exhaustively over a bounded model, kept honest by trace conformance and shown to be checkable
by construction, and the self-improvement loop is bound by construction and can only tighten itself. We then ran
the same method forward through six more releases and reported what happened to the guarantees. The action-safety
core held at every release, one release added capability and needed no new invariant, and six new machine-checked
families carried the core outward into memory, agency, epistemic honesty, a harness that runs many governed agents
on one thread, the self-improvement loop that changes the system itself, and the residency of what that loop
produces, each with the broken models that show its checker can fail. The last of those is also the first we take
onto real hardware, where gated self-improvement compounds a real model while rejecting the changes that only look
like gains. We defined the exact scope of each claim and published the runtime, tools, and gate suite so that the
results are reproducible and the gates can be run against other systems. The part meant to outlast this particular
runtime was always the discipline. This paper is the evidence that it lasts.

\section*{Data and code availability}

The runtime (\texttt{antahkarana}~\citep{antahkarana_sdk}), the zero-dependency operator tool
(\texttt{antahkarana-cli}~\citep{antahkarana_cli}), the full 563-test gate suite, the run artifacts, the per-rung
checkers, and the canonical TLA+ specification are open. Every figure regenerates from the run data with a single
script, and the central non-bypass result reproduces with one command (\texttt{atk verify -{}-teeth}). The broader
design philosophy is set out in the book \emph{\antah{}: The Inner Instrument}~\citep{soni2026innerinstrument}.

\begin{itemize}[leftmargin=1.4em,itemsep=2pt]
\item Runtime (PyPI): \url{https://pypi.org/project/antahkarana/}
\item Operator CLI (PyPI): \url{https://pypi.org/project/antahkarana-cli/}
\item Model card (Hugging Face): \url{https://huggingface.co/deepakdsoni/antahkarana-base}
\item Book (Apple Books): \url{https://books.apple.com/us/book/id6785227569}
\end{itemize}

\bibliographystyle{plainnat}
\bibliography{references}

\end{document}